\documentclass[%
nofootinbib,
superscriptaddress,
preprint,
 amsmath,amssymb,ams
 aps,
]{revtex4-1}

\usepackage{graphicx}% Include figure files
\usepackage{dcolumn}% align table columns on decimal point
\usepackage{bm}% bold math
\usepackage{multirow}
\usepackage{tikz}
\usepackage{slashed}
\usetikzlibrary{shapes.misc}

\begin{document}

\preprint{KEK-TH-2116}

\title{Renormalization of Unitarized Weinberg-Tomozawa Interaction without On-shell Factorization \\  and
$I=0$ $\bar K N$ - $\pi\Sigma$ Coupled Channels}% Force line breaks with \\

\author{Osamu Morimatsu}
\affiliation{%
 Theory Center, Institute of Particle and Nuclear Studies (IPNS), \\
High Energy Accelerator Research Organization (KEK), \\
1-1 Oho, Tsukuba, Ibaraki, 205-0801, Japan}%
\affiliation{Department of Physics, Faculty of Science, University of Tokyo, \\
7-3-1 Hongo Bunkyo-ku Tokyo 113-0033, Japan}%Lines break automatically or can be forced with \\
\affiliation{%
Department of Particle and Nuclear Studies, \\
Graduate University for Advanced Studies (SOKENDAI), \\
1-1 Oho, Tsukuba, Ibaraki 305-0801, Japan
}%
\author{Kazuki Yamada}%
\affiliation{%
 Theory Center, Institute of Particle and Nuclear Studies (IPNS), \\
High Energy Accelerator Research Organization (KEK), \\
1-1 Oho, Tsukuba, Ibaraki, 205-0801, Japan}%
\affiliation{Department of Physics, Faculty of Science, University of Tokyo, \\
7-3-1 Hongo Bunkyo-ku Tokyo 113-0033, Japan}%Lines break automatically or can be forced with \\
 \email{osamu.morimatsu@kek.jp}
 \email{Second.Author@institution.edu}

\date{\today}% It is always \today, today,

\begin{abstract}
We calculate the scattering $T$-matrix of $I=0$ $\bar K N-\pi \Sigma$ coupled channels taking a ladder sum of the Weinberg-Tomozawa interaction without on-shell factorization,
regularizing three types of divergent meson-baryon loop functions by dimensional regularization and renormalizing them by introducing counter terms.
We show that not only infinite but also finite renormalization is important in order for the renormalized physical scattering $T$-matrix to have the form of the Weinberg-Tomozawa interaction.
The results with and without on-shell factorization are compared.
The difference of the  scattering $T$-matrix is small near the renormalization point, close to the observed $\Lambda$(1405).
The difference, however, increases with the distance from the renormalization point.
The scattering $T$-matrix without on-shell factorization has two poles in the complex center-of-mass energy plane as with on-shell factorization, the real part of which is close to the observed $\Lambda$(1405).
While the difference is small with and without on-shell factorization in the position of the first pole, closer to the observed $\Lambda$(1405),
the difference is considerably large in the position of the second pole:
the imaginary part of the center-of-mass energy of the second pole without on-shell factorization is as large as or even larger than twice that with on-shell factorization.
Also, we discuss the origin of the contradiction about the second pole between two approaches, the chiral unitary approach with on-shell factorization and the phenomenological approach without on-shell factorization.
\begin{description}
\item[PACS numbers]
13.75.Jz, 14.20.-c, 11.30.Rd
\end{description}
\end{abstract}

\pacs{Valid PACS appear here}
\maketitle
\section{Introduction}
Chiral perturbation theory \cite{Weinberg:1978kz,Gasser:1983yg,Gasser:1984gg,Scherer:2009bt} is a method to describe the dynamics of Goldstone bosons in the framework of an effective field theory.
Writing down the most general effective Lagrangian containing all possible terms compatible with chiral symmetry, one obtains the scattering $T$-matrix order by order in powers of momenta and quark masses at low center-of-mass energies, where infinities arising from loops are absorbed in a renormalization of the coefficients of the effective Lagrangian.
Chiral perturbation theory has been successful in describing low-energy meson-meson and meson-baryon scatterings but cannot describe bound states or resonances due to its very perturbative nature.

A nonperturbative method, the chiral unitary approach has been developed \cite{Kaiser:1995eg,Oset:1997it,Oller:2000fj,Lutz:2001yb}, in which the leading terms of chiral perturbation are resumed by means of integral equations, such as the Lippman-Schwinger equation or dispersion relations, the $N/D$ method.
The chiral unitary approach sacrifices the systematics of chiral perturbation theory but accommodates bound states or resonances.

One of the applications of the chiral unitary approach, which have received much attention in the past decades, is the $\Lambda(1405)$ \cite{Oller:2000fj,Lutz:2001yb,Jido:2003cb,Magas:2005vu,Hyodo:2007jq,Hyodo:2011ur,Ikeda:2011pi,Ikeda:2012au}. 
In the chiral unitary approach the scattering $T$-matrix analytically continued in the complex center-of-mass energy plane turns out to have two poles close to the observed $\Lambda(1405)$, both contributing to the final experimental invariant mass distribution.
It should be noted, however, that they employed an approximation, on-shell factorization, which approximates the off-shell interaction vertex by the on-shell interaction vertex and takes out from the meson-baryon loop integral.

Recently, this double-pole interpretation of the $\Lambda(1405)$ has been questioned {\cite{Akaishi:2010wt,Revai:2017isg,Myint:2018ypc}.
In particular, in Ref.\ {\cite{Revai:2017isg}}, it was claimed that the energy dependence of the chiral based $\bar{K}N$ potentials, responsible for the
occurrence of two poles in the $I = 0$ sector, is the consequence of applying on-shell factorization.
When the dynamical equation is solved without on-shell factorization, the scattering $T$-matrix has only one pole in the complex center-of-mass energy plane, close to the observed $\Lambda(1405)$.
The argument, however, is based on a nonrelativistic phenomenological potential model, a separable potential model, with cut-off functions.
Therefore, it is not clear whether the contradiction between two approaches is due to the difference in the approximation, with or without on-shell factorization,
or due to the difference in the theoretical framework, chiral interaction with relativistic kinematics or phenomenological interaction with nonrelativistic kinematics.

The purpose of the present paper is as follows.
First, we would like to show that by renormalizing the divergent loop terms we can calculate the meson-baryon scattering $T$-matrix in the chiral unitary approach without employing on-shell factorization.
Then, we would like to see whether the second pole is found in the complex center-of-mass energy plane near the observed $\Lambda(1405)$.
Consequently, we would like to clarify the origin of the contradiction about the second pole for the $\Lambda(1405)$ between two approaches, the chiral unitary approach with on-shell factorization and the phenomenological approach without on-shell factorization.

\section{Formulation}
The Weinberg-Tomozawa interaction, the lowest-order term of chiral perturbation in the meson-baryon channel, is given by
\begin{align}
  {\cal L}&= i\frac{C}{4f^2} \bar B' M' \overleftrightarrow{\slashed \partial} M B + \text{counter terms},
\end{align}
where $B$ and $B'$ are baryon fields and $M$ and $M'$ are meson (Goldstone boson) fields.
\subsection{single-channel}
Let us first consider a single-channel scattering of a meson $M$ and a baryon $B$, $M(k) + B(p) \rightarrow M(k') + B(p')$, where $k$ ($k'$) and $p$ ($p'$) are four momenta of the incoming (outgoing) meson and baryon, respectively.    
The scattering $T$-matrix of the renormalized ladder sum of the Weinberg-Tomozawa interaction, $T$, is given by Fig.\ \ref{fig.1},
\begin{align}
  T &= T_{tree} + \left( T^{bare}_{one \mathchar`- loop} + \delta T_{one \mathchar`- loop} \right) + \cdots \nonumber \\
  &= T_{tree} + T_{one \mathchar`- loop} + \cdots,
\end{align}
where $T_{tree}$ is the tree term, $T^{bare}_{one \mathchar`- loop}$ and $\delta T_{one \mathchar`- loop}$ are
the bare one-loop term and its counter term, respectively, $T_{one \mathchar`- loop}$ is the renormalized
one-loop term, i.e.\ the sum of $T^{bare}_{one \mathchar`- loop}$ and $\delta T_{one \mathchar`- loop}$, and $\cdots$
represents higher loop terms.
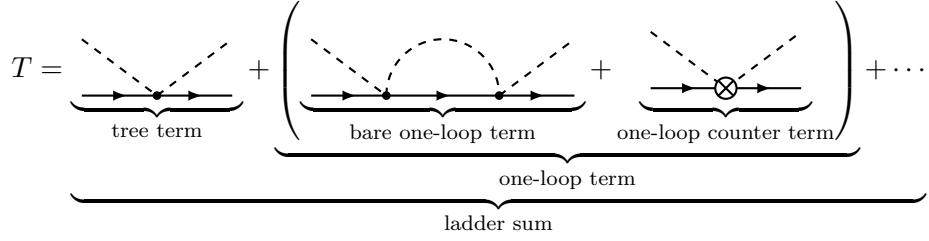
\begin{figure}[h]
\[
T=
\underbrace{
\underbrace{
\raisebox{-0.3cm}
{
\begin{tikzpicture}[>=latex]
  \draw[thick,->]       (0,0) -- (0.6,0);
  \draw[thick]        (0.5,0) -- (1,0);
  \draw[thick,->]        (1,0) -- (1.6,0);
  \draw[thick]        (1.5,0) -- (2,0);
  \draw [dashed, thick] (0,0.75) -- (1,0);
  \draw [dashed, thick] (1,0) -- (2,0.75);
  \draw [fill] (1,0) circle [radius=0.05];
\end{tikzpicture}
}
}_\text{tree term}
+
\underbrace{
\left(
\underbrace{
\raisebox{-0.3cm}
{
\begin{tikzpicture}[>=latex]
  \draw[thick,->]        (0,0) -- (0.6,0);
  \draw[thick]        (0.5,0) -- (1,0);
  \draw[thick,->]        (1,0) -- (1.85,0);
  \draw[thick]        (1.75,0) -- (2.5,0);
  \draw[thick,->]        (2.5,0) -- (3.1,0);
  \draw[thick]        (3.0,0) -- (3.5,0);
  \draw [dashed, thick] (0,0.75) -- (1,0);
  \draw [dashed, thick] (1,0) arc [radius=0.75, start angle=180, end angle= 0];
  \draw [dashed, thick] (2.5,0) -- (3.5,0.75);
  \draw [fill] (1,0) circle [radius=0.05];
  \draw [fill] (2.5,0) circle [radius=0.05];
\end{tikzpicture}
}
}_\text{bare one-loop term}
+
\underbrace{
\raisebox{-0.3cm}
{
\begin{tikzpicture}[>=latex]
  \draw[thick,->]       (0,0) -- (0.6,0);
  \draw[thick]        (0.5,0) -- (1,0);
  \draw[thick,->]        (1,0) -- (1.6,0);
  \draw[thick]        (1.5,0) -- (2,0);
  \draw [dashed, thick] (0,0.75) -- (1,0);
  \draw [dashed, thick] (1,0) -- (2,0.75);
  \draw [fill=white,thick] (1,0) circle [radius=0.1414];
  \draw [thick] (0.9,0.1) -- (1.1,-0.1);
  \draw [thick] (0.9,-0.1) -- (1.1,0.1);
\end{tikzpicture}
}
}_\text{one-loop counter term}
\right)
}_\text{one-loop term}
+ \cdots
}_\text{ladder sum}
\]
\caption{\label{fig.1} Diagrammatic representation of meson-baryon scattering $T$-matrix in the ladder sum.}
\end{figure}
The bare one-loop crossed term in Fig.\ \ref{fig.2} is not taken into account, so that crossing symmetry is broken in the scattering $T$-matrix of ladder sum, Eq.\ (2).
\begin{figure}[h]
\begin{center}
\begin{tikzpicture}[>=latex]
  \draw[thick,->]        (0,0) -- (0.6,0);
  \draw[thick]        (0.5,0) -- (1,0);
  \draw[thick,->]        (1,0) -- (1.85,0);
  \draw[thick]        (1.75,0) -- (2.5,0);
  \draw[thick,->]        (2.5,0) -- (3.1,0);
  \draw[thick]        (3.0,0) -- (3.5,0);
  \draw [dashed, thick] (0,0.75) -- (2.5,0);
  \draw [dashed, thick] (1,0) arc [radius=0.75, start angle=180, end angle= 0];
  \draw [dashed, thick] (1.0,0) -- (3.5,0.75);
  \draw [fill] (1,0) circle [radius=0.05];
  \draw [fill] (2.5,0) circle [radius=0.05];
\end{tikzpicture}
\caption{\label{fig.2} The bare one-loop crossed diagram.}
\end{center}
\end{figure}
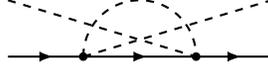

The tree term is given by
\begin{align}
  T_{tree} = & \left( - \frac{C}{4f^2} \right) \bar u(p') \left(\slashed {k} + \slashed {k}'\right) u(p) \nonumber \\
  = & - \frac{C}{4f^2} \bar u(p') 2(\slashed P - M) u(p),
\end{align}
where $P$ is the total momentum of the system, $P=p+k=p'+k'$ and $u(p)$ ($\bar u(p')$) is the Dirac spinor for the incoming (outgoing) baryon.
The bare one-loop term is given by
\begin{align}
  T^{bare}_{one \mathchar`- loop} = & \left(- \frac{C}{4f^2}\right)^2 \bar u(p') i\int\frac{d^4q}{(2\pi)^4} \left(\slashed {q} + \slashed {k}'\right) \frac{2M} {\left[(P-q)^2-M^2\right](q^2-m^2)} \left(\slashed {k} + \slashed {q}\right) u(p) \nonumber \\
  = & \left(- \frac{C}{4f^2}\right)^2 \bar u(p') \left[ G_0 + G_1 \slashed {k}' \slashed {P} + G_1 \slashed {P} \slashed {k} + G_2 \slashed {k}' \slashed {k} \right] u(p) \nonumber \\
  = & \left(- \frac{C}{4f^2}\right)^2 \bar u(p') \left[ G_0 + G_1 M \left( \slashed {k} + \slashed {k}'\right) + \left( 2G_1 + G_2 \right) \slashed {k}' \slashed {k} \right] u(p) \nonumber \\
  = & \left(- \frac{C}{4f^2}\right)^2 \bar u(p') \left[ G_0 + G_1 2 M \left(\slashed P-M\right) + \left( 2G_1 + G_2 \right) \left(\slashed P-M\right)^2 \right] u(p),
\end{align}
where
\begin{align}
\left\{
  \begin{aligned}
  &i\int\frac{d^4q}{(2\pi)^4} \frac{2Mq^2} {\left[(P-q)^2-M^2\right](q^2-m^2)} \equiv G_0 \\
  &i\int\frac{d^4q}{(2\pi)^4} \frac{2M\slashed {q}} {\left[(P-q)^2-M^2\right](q^2-m^2)} \equiv G_1 \slashed{P} \\
  &i\int\frac{d^4q}{(2\pi)^4} \frac{2M} {\left[(P-q)^2-M^2\right](q^2-m^2)} \equiv G_2.
  \end{aligned}
\right.
\end{align}
The Klein-Gordon propagator is employed not only for the meson but also for the baryon for comparison,
because the calculations in the chiral unitary approach with on-shell factorization, Ref.\ \cite{Oller:2000fj,Jido:2003cb,Hyodo:2007jq,Hyodo:2011ur}, are regarded as to employ the Klein-Gordon propagator for the baryon, though it is explained that the $N/D$ method is used.
It is, however, not difficult to employ the Dirac propagator instead of the Klein-Gordon propagator for the baryon.   

 $G_0$ is quadratically divergent while $G_1$ and $G_2$ are logarithmically divergent, which are given in dimensional regularization as
\begin{align}
&\left\{
  \begin{aligned}
  G_0 =&  \frac{2M}{16\pi^2} \left[ (M^2+m^2) \left( -\frac{2}{\epsilon} + \gamma - \log{4\pi} \right) -\frac{1}{6}(P^2-3M^2-3m^2)
  + \int_0^1 dx (2\Delta+x^2P^2) \log \Delta \right] \\
  G_1 =& \frac{2M}{16\pi^2} \left[ \frac{1}{2} \left( - \frac{2}{\epsilon} + \gamma - \log{4\pi} \right) + \int_0^1 dx x \log \Delta \right] \\
  G_2 =& \frac{2M}{16\pi^2} \left[ - \frac{2}{\epsilon} + \gamma - \log{4\pi} + \int_0^1 dx \log \Delta \right],
  \end{aligned}
\right.
\end{align}
where $\Delta = xM^2+(1-x)m^2-x(1-x)P^2$, $\epsilon=d-4$, $d$ is the dimension of space and time and $\gamma$ is the Euler constant.

When $G_0$, $G_1$ and $G_2$ are Taylor expanded in $P^2-M^2$ as
\begin{align}
  G_i=G^{(0)}_i+G'_i{}^{(0)} (P^2-M^2) + \frac{1}{2} G''_i{}^{(0)} \left( P^2-M^2 \right)^2 + \cdots,
\end{align}
where the divergences appear only in $G^{(0)}_0$, $G^{(0)}_1$ and $G^{(0)}_2$, the zeroth order
coefficients of $G_0$, $G_1$ and $G_2$, respectively.
Then, $T^{bare}_{one \mathchar`- loop}$ is Taylor expanded in $\slashed P-M$ as
\begin{align}
  T^{bare}_{one \mathchar`- loop} = & \left(-\frac{C}{4f^2}\right)^2 \bar u(p')  \left[ G^{(0)}_0
  + \left( G'^{(0)}_0 + G^{(0)}_1 \right) 2M \left( \slashed P-M \right) \right. \nonumber \\
  &  + \left\{ \left( G'^{(0)}_0 + 2 G^{(0)}_1 + G^{(0)}_2 \right)  +  \left( G''^{(0)}_0 + 2 G'^{(0)}_1\right) 2M^2 \right\} \left( \slashed P-M \right)^2 \nonumber \\
  &  \left. + {\cal O} \left( \left( \slashed P-M \right)^3 \right) \right] u(p),
\end{align}
where divergences appear in the coefficients of $1$, $\slashed P-M$ and $\left(\slashed P-M\right)^2$.
Therefore, we need three counter terms proportional to $1$, $\slashed {k} + \slashed {k}'$ and $\slashed {k} \slashed {k}'$ in order to cancel divergences in $T^{bare}_{one \mathchar`- loop}$:
\begin{align}
  \delta T_{one-loop}
  = & \bar u(p')  \left[ \delta_0 + \delta_1 \left(\slashed {k} + \slashed {k}'\right) + \delta_2 \slashed {k} \slashed {k}' \right] u(p) \nonumber \\
  = & \bar u(p')  \left[ \delta_0 + \delta_1 2\left( \slashed P-M \right) + \delta_2 \left( \slashed P-M \right)^2  \right] u(p).
\end{align}
The origin of these terms in the context of the effective field theory will be discussed elsewhere.
We determine finite terms in $T_{one \mathchar`- loop}$ by requiring that
$T_{tree} + T_{one \mathchar`- loop}$ is the same as $T_{tree}$ up to ${\cal O}\left(\left(\slashed P-M\right)^2\right)$:
\begin{align}
  T_{tree} + T_{one \mathchar`- loop} = &- \frac{C}{4f^2} \bar u(p') \left(\slashed {k} + \slashed {k}'\right) u(p) + {\cal O}\left( \left( \slashed P-M \right)^2 \right),%\\
\end{align}
which gives
\begin{align}
\left\{
  \begin{aligned}
  &\left(-\frac{C}{4f^2}\right)^2 G^{(0)}_0 + \delta_0 = 0 \\
  &\left(-\frac{C}{4f^2}\right)^2 \left( G'^{(0)}_0 + G^{(0)}_1 \right) 2M + 2\delta_1 = 0 \\
  &\left(-\frac{C}{4f^2}\right)^2 \left( G'^{(0)}_0 + 2 G^{(0)}_1 + G^{(0)}_2 \right) + \delta_2 = \text{finite constant},
  \end{aligned}
\right.
\end{align}
where \lq finite constant' is not determined by the above requirement and will be discussed later.
We define finite renormalized loop functions, $G^R_0$, $G^R_1$ and $G^R_2$ by
\begin{align}
\left\{
  \begin{aligned}
  &G^R_0 = G_0 - G^{(0)}_0 \\
  &G^R_1 = G_1 - G^{(0)}_1-G'_0{}^{(0)} \\
  &G^R_2 = G_2 - G^{(0)}_2 + \text{finite constant}.
  \end{aligned}
\right.
\end{align}
$G^R_0$, $G^R_1$ and $G^R_2$ are expressed as
\begin{align}
\left\{
  \begin{aligned}
  G^R_0 =&  \frac{2M}{16\pi^2} \left[ (M^2+m^2) \left(a_0 (\mu) + 2 - \log\mu^2\right) -\frac{1}{6}(P^2-3M^2-3m^2) 
  + \int_0^1 dx (2\Delta+x^2P^2) \log \Delta \right] \\
  G^R_1 =& \frac{2M}{16\pi^2} \left[ \frac{1}{2}\left(a_1 (\mu) + 2 - \log\mu^2\right) + \int_0^1 dx x \log \Delta \right] \\
  G^R_2 =& \frac{2M}{16\pi^2} \left[ a_2 (\mu) + 2 - \log\mu^2 + \int_0^1 dx \log \Delta \right].
  \end{aligned}
\right.
\end{align}
$\mu$ is the renormalization scale and $a_0$, $a_1$ and $a_2$ are subtraction constants, which are determined to satisfy
\begin{align}
\left\{
  \begin{aligned}
  &G^R_0{}^{(0)} = 0 \\
  &G^R_0{}'^{(0)} + G^R_1{}^{(0)} = 0 \\
  &G^R_2{}^{(0)} = \text{finite constant}.
  \end{aligned}
\right.
\end{align}
Carrying out integrals in Eq.\ (13) we obtain explicit expressions for $G^R_0$, $G^R_1$ and $G^R_2$ as
\begin{align}
\left\{
  \begin{aligned}
  &G^R_0 =\frac{2M} {16\pi^2} \left\{ \left(M^2+m^2\right)\left(a_0(\mu)+ \log \frac{M^2}{\mu^2}\right)+\frac{3}{2}s+M^2-2{\bar q}^2+\frac{3(M^2-m^2)^2+4s^2}{2s}\right.  \\
  &\qquad \left.+ \frac{M^2-m^2-s}{2s}\frac{(M^2-m^2-s)^2-4s(\bar q^2+2m^2)}{4s}\log\frac{m^2}{M^2} \right. \\
  &\qquad \left.+ \frac{\bar q}{\sqrt{s}}\frac{5(M^2-m^2-s)^2+4s(\bar q^2+2m^2)}{4s}\log \frac{\phi_{++}\phi_{+-}}{\phi_{-+}\phi_{--}}  \right\}, \\
  &G^R_1 = \frac{M}{16\pi^2}\left\{ a_1(\mu)+\log \frac{M^2}{\mu^2}+\frac{M^2-m^2}{s}+\frac{(M^2-m^2-s)^2+4 s{\bar q}^2}{4s^2}\log\frac{m^2}{M^2}\right. \\
  &\qquad \left.+\frac{\bar q}{\sqrt{s}}\frac{M^2-m^2-s}{2s} \log \frac{\phi_{++}\phi_{+-}}{\phi_{-+}\phi_{--}}  \right\} \\
  &G^R_2 = \frac{2M}{16\pi^2} \left\{ a_2(\mu) + \log \frac{M^2}{\mu^2} + \frac{m^2-M^2+s}{2s} \log \frac{m^2}{M^2}
  + \frac{\bar q}{\sqrt{s}} \log \frac{\phi_{++}\phi_{+-}}{\phi_{-+}\phi_{--}} \right\},
  \end{aligned}
\right.
\end{align}
where $s=P^2$,
$\bar q = \sqrt{(s-(M-m)^2)(s-(M+m)^2)}/(2\sqrt{s})$
and
$\phi_{\pm\pm} = \pm s \pm (M^2-m^2)+2\bar q\sqrt{s}$.
Then, $T_{one \mathchar`- loop}$ is given in terms of the finite renormalized loop functions as
\begin{align}
  T_{one \mathchar`- loop} =& \left(- \frac{C}{4f^2}\right)^2 \bar u(p') \left[ G^R_0 + G^R_1 \slashed {k}' \slashed {P} + G^R_1 \slashed {P} \slashed {k} + G^R_2 \slashed {k}' \slashed {k} \right] u(p) \nonumber \\
  =& \left(- \frac{C}{4f^2}\right)^2 \bar u(p')
  \begin{pmatrix}
    \slashed{k'} & 1
  \end{pmatrix}
  \hat G^R
  \begin{pmatrix}
    1 \\
    \slashed{k}
  \end{pmatrix}
  u(p),
 \end{align}
where $\hat G^R$ is a 2 by 2 matrix defined by
\begin{align}
  \hat G^R =
  \begin{pmatrix}
    G^R_1 \slashed {P} & G^R_2 \\
    G^R_0 & G^R_1  \slashed {P}
  \end{pmatrix}.
 \end{align}
Then, summing up the ladder terms we can express $T$ in terms of the renormalized loop functions as
\begin{align}
  T = & \bar u(p')
  \begin{pmatrix}
    \slashed{k'} & 1
  \end{pmatrix}
  \left\{
  - \frac{C}{4f^2} + \left( - \frac{C}{4f^2}\right)^2 \hat G^R + \left( - \frac{C}{4f^2}\right)^3 \left (\hat G^R \right)^2 + \cdots \right\}
  \begin{pmatrix}
    1 \\
    \slashed{k}
  \end{pmatrix}
   u(p) \nonumber \\
  = & \bar u(p')
  \begin{pmatrix}
    \slashed{k'} & 1
  \end{pmatrix}
  \frac{- \frac{C}{4f^2}}{1 - \left(- \frac{C}{4f^2}\right) \hat G^R}
  \begin{pmatrix}
    1 \\
    \slashed{k}
  \end{pmatrix}
   u(p) \nonumber \\
   = & 
  \bar u(p') \frac{ - \frac{C}{4f^2} \left(\slashed {k} + \slashed {k}'\right)
  + \left(- \frac{C}{4f^2}\right)^{2} \left( G^R_0 - G^R_1 \slashed {k}' \slashed {P} - G^R_1 \slashed {P} \slashed {k} + G^R_2 \slashed {k}' \slashed {k} \right)
  }
  {\left\{1 - \left(- \frac{C}{4f^2}\right) G^R_1 \slashed {P} \right\}^2 - \left(- \frac{C}{4f^2}\right)^2 G^R_0G^R_2} 
   u(p).
\end{align}
Expanding in $\slashed P-M$ and using Eq.\ (14), one can show that 
\begin{align}
  T = &- \frac{C}{4f^2} \bar u(p') \left(\slashed {k} + \slashed {k}'\right) u(p) + {\cal O}\left( \left( \slashed P-M \right)^2 \right).
\end{align}
Namely, once the one-loop term is properly renormalized, no further renormalization, neither infinite nor finite renormalization, is necessary for the scattering $T$-matrix in the ladder sum.

\subsection{coupled channels}
Let us move on to a meson-baryon scattering of coupled $n$-channels.
We introduce $2n$ by $2n$ matrices, ${\boldsymbol \Lambda}$ and ${\boldsymbol G}^R$, with both channel indices $i$, $j$ and the index of the 2 by 2 matrix, which already appeared in the single-channel scattering as in Eq.\ (17).
\footnote{$2n$ by $2n$ matrix representation of the scattering equation has been presented also in Ref.\ \cite{Revai:2017isg}.}
For given channel indices, $i$ and $j$, $\left[\boldsymbol\Lambda\right]_{ji}$ and $\left[\hat {\boldsymbol G}^R \right]_{ji}$ are defined to be 2 by 2 matrices as
\begin{align}
  \left[\boldsymbol\Lambda\right]_{ji} 
  = \frac{C_{ji}}{4f^2} {\boldsymbol 1} 
  =
  \begin{pmatrix}
  \frac{C_{ji}}{4f^2} & 0 \\
  0 & \frac{C_{ji}}{4f^2} \\
  \end{pmatrix},
\end{align}
and
\begin{align}
  \left[\hat {\boldsymbol G}^R \right]_{ji} 
  = \delta_{ji} \hat G^R_i 
  = \delta_{ji}
  \begin{pmatrix}
    G^R_{i1} \slashed {P} & G^R_{i2} \\
    G^R_{i0} & G^R_{i1}  \slashed {P}
  \end{pmatrix}.
\end{align}
Namely, $\boldsymbol\Lambda$ and $\hat {\boldsymbol G}^R$ are diagonal with respect to indices of 2 by 2 matrices and channel indices, respectively.

We impose the same renormalization conditions as in the single-channel scattering, Eq.\ (14), for the renormalized loop functions in each channel:
\begin{align}
\left\{
  \begin{aligned}
  &G^R_{i0}{}^{(0)} = 0 \\
  &G^R_{i0}{}'^{(0)} + G^R_{i1}{}^{(0)} = 0 \\
  &G^R_{i2}{}^{(0)} = \text{finite constant}.
  \end{aligned}
\right.
\end{align}
The scattering $T$-matrix from the channel $i$ to the channel $j$ is given by
\begin{align}
  T_{ji} = & \bar u_j(p'_j)
  \begin{pmatrix}
    \slashed{k'_j} & 1
  \end{pmatrix}
  \left[ -{\boldsymbol\Lambda} + (-{\boldsymbol\Lambda}) \hat {\boldsymbol G}^R (-{\boldsymbol\Lambda}) + \cdots \right]_{ji}
  \begin{pmatrix}
    1 \\
    \slashed{k_i}
  \end{pmatrix}
   u_i(p_i) \nonumber \\
 = & \bar u_j(p'_j)
  \begin{pmatrix}
    \slashed{k'_j} & 1
  \end{pmatrix}
  \left[ {-{\boldsymbol\Lambda}} \left({\boldsymbol 1} - \hat {\boldsymbol G}^R \left(- {\boldsymbol\Lambda}\right)\right)^{-1} \right]_{ji}
  \begin{pmatrix}
    1 \\
    \slashed{k_i}
  \end{pmatrix}
   u_i(p_i).
\end{align}
Eq.\ (23) together with Eq.\ (22) is the main result in the formulation section of the present paper.

\subsection{On-shell factorization}
Here, we summarize minimum expressions for the scattering $T$-matrix of the ladder sum of the Weinberg-Tomozawa interaction with on-shell factorization
because we compare the results with and without on-shell factorization.

In a single-channel meson-baryon scattering, the tree term is given irrespective of on-shell factorization as the matrix element of the on-shell interaction vertex as
\begin{align}
  T_{tree} = & \bar u(p') \left(- \frac{C}{4f^2}\right) \left(\slashed {k} + \slashed {k}'\right) u(p) \nonumber \\
  \approx & \chi'{}^\dagger \left(- \frac{C}{4f^2}\right) 2(\sqrt{s} - M) \frac{E+M}{2M} \chi \nonumber \\
  \equiv & \chi'{}^\dagger V_{on} \chi,
\end{align}
where $\chi$ and $\chi'^\dagger$ are Pauli spinors for the incoming and outgoing baryons, respectively.
Then, in the bare one-loop term, $T^{bare}_{one \mathchar`- loop}$, the off-shell interaction vertex is approximated by the on-shell interaction vertex and is taken out from the loop integral as
\begin{align}
  T^{bare}_{one \mathchar`- loop} = &  \bar u(p') i\int\frac{d^4q}{(2\pi)^4} \left(- \frac{C}{4f^2}\right) \left(\slashed {q} + \slashed {k}'\right) \frac{2M} {\left[(P-q)^2-M^2\right](q^2-m^2)} \left(- \frac{C}{4f^2}\right) \left(\slashed {k} + \slashed {q}\right) u(p) \nonumber \\
  \rightarrow & \chi'{}^\dagger V_{on} \left\{ i\int\frac{d^4q}{(2\pi)^4} \frac{2M} {\left[(P-q)^2-M^2\right](q^2-m^2)} \right\} V_{on} \chi \nonumber \\
  = & \chi'{}^\dagger V_{on} G V_{on} \chi,
\end{align}
where $G$ is nothing but $G_2$ in Eq.\ (5).
In Ref.\ \cite{Oller:2000fj,Jido:2003cb,Hyodo:2007jq,Hyodo:2011ur} it is explained that the finite unitary scattering $T$-matrix is obtained by dispersion relations, the $N/D$ method, and renormalization is not explicitly mentioned.
It is, however, equivalent to renormalize the scattering $T$-matrix by introducing the counter term,
\begin{align}
  \delta T_{one-loop}
  = & \chi'{}^\dagger  \delta \left( \sqrt{s}-M \right)^2 \left(\frac{E+M}{2M}\right)^2 \chi,
\end{align}
which corresponds to the term with $\delta_2$ in Eq.\ (9).
The terms with $\delta_0$ and $\delta_1$ in Eq.\ (9) do not appear in on-shell factorization. 
The renormalized one-loop term, $T_{one \mathchar`- loop}$, i.e.\ the sum of $T^{bare}_{one \mathchar`- loop}$
and $\delta T_{one \mathchar`- loop}$, is given by
\begin{align}
  T_{one \mathchar`- loop} = \chi'{}^\dagger V_{on} G^R V_{on} \chi,
\end{align}
where $G^R$ is $G^R_2$ in Eq.\ (12).
Then, the ladder sum is
\begin{align}
  T &= \chi'{}^\dagger \left( V_{on} + V_{on} G^R V_{on} + V_{on} G^R V_{on} G^R V_{on} + \cdots \right) \chi \nonumber \\
  &= \chi'{}^\dagger \frac{V_{on}}{1 - V_{on} G^R} \chi,
\end{align}
which, by the use of Eq.\ (23), becomes,
\begin{align}
  T &= \chi'{}^\dagger \frac{- \frac{C}{4f^2} 2(\sqrt{s} - M) \frac{E+M}{2M}}{1 -\left(- \frac{C}{4f^2} \right) 2(\sqrt{s} - M) \frac{E+M}{2M} G^R} \chi \nonumber \\
  &= - \chi'{}^\dagger \frac{C}{4f^2} 2(\sqrt{s} - M) \frac{E+M}{2M} \chi + {\cal O}((\sqrt{s} - M)^2).
\end{align}
Namely, without renormalization the scattering $T$-matrix in the ladder sum is the same as $T_{tree}$ up to ${\cal O}\left(\left(\sqrt{s}-M\right)^2\right)$ in on-shell factorization;
on-shell factorization obscures the importance of renormalization.

In a meson-baryon scattering of coupled $n$-channels, the scattering $T$-matrix from the channel $i$ to the channel $j$ is given by
\begin{align}
  T_{ji} = & \chi_j^\dagger
  \left[ {\boldsymbol V_{on}} + {\boldsymbol V_{on}} {\boldsymbol G^R} {\boldsymbol V_{on}} + \cdots \right]_{ji}
   \chi_i \nonumber \\
 = & \chi_j^\dagger
  \left[ {\boldsymbol V_{on}} \left({\boldsymbol 1} - {\boldsymbol G}^R {\boldsymbol V_{on}}\right)^{-1} \right]_{ji}
   \chi_i,
\end{align}
where $\boldsymbol V_{on}$ and ${\boldsymbol G}^R$ are $n$ by $n$ matrices with channel indices,
\begin{align}
  \left[\boldsymbol V_{on}\right]_{ji} 
  = -\frac{C_{ji}}{4f^2} \left( 2\sqrt{s} - M_i - M_j \right) \sqrt{\frac{E_j+M_j}{2M_j}} \sqrt{\frac{E_i+M_i}{2M_i}},
\end{align}
and
\begin{align}
  \left[\hat {\boldsymbol G}^R \right]_{ji} 
  = \delta_{ji} G^R_i.
\end{align}

\section{results and discussion}
Now, we compare the results of the calculation with and without on-shell factorization.
In the calculation of the chiral unitary approach with on-shell factorization, nonrelativistic approximation, Eq.\ (24), has been adopted for the matrix elements with respect to Dirac spinors. Hereafter, we adopt the same approximation in our calculation for comparison.
In the chiral unitary approach with on-shell factorization it was shown in Ref.\ \cite{Hyodo:2007jq} that the results of the $\pi \Sigma-{\bar K} N$ coupled channels and the $\pi \Sigma - {\bar K} N - \eta \Lambda - K \Xi$ coupled channels are nearly the same.
Therefore, we calculate the scattering $T$-matrix of the $\pi \Sigma - {\bar K} N$ coupled channels for simplicity and compare the results with those of Ref.\ \cite{Hyodo:2007jq},
where the parameters $f$ and $\mu$ are taken to be the same as in Ref.\ \cite{Hyodo:2007jq}, i.e.\ $f = 106.95$ MeV and $\mu = 630$ MeV.
As we have already mentioned, we require that the scattering $T$-matrix has the form of the Weinberg-Tomozawa interaction at $\sqrt{s} = M$.
While this determines the subtraction constants $a_0$ and $a_1$, we need another condition for $a_2$.
We adopt the following two cases of the condition and check how the results depend on them.
One, case $A$, is that the second-order derivative of the single-channel scattering $T$-matrix is the same as that of the on-shell factorization at $\sqrt{s} = M$, and
the other, case $B$, is that the single-channel scattering $T$-matrix is the same as that of the on-shell factorization at $\sqrt{s} = M+m$,
\begin{align}
  &A: \qquad \left.\frac{\partial^2 T}{\partial\sqrt{s}^2}\right|_{\sqrt{s}=M} = \left.\frac{\partial^2 T^{on-shell}}{\partial\sqrt{s}^2}\right|_{\sqrt{s}=M},  \\
  &B: \qquad \left.T\right|_{\sqrt{s}=M+m} = \left.T^{on-shell}\right|_{\sqrt{s}=M+m}.
\end{align}
The subtraction constants $a_0$, $a_1$ and $a_2$ for cases $A$ and $B$ together with $a_2$ in on-shell factorization, case $C$, are summarized in Table I.
\begin{table}[h]
\begin{tabular}{p{1.5cm}|p{2cm}p{2cm}p{2cm}|p{2cm}p{2cm}p{2cm}}
\hline
\hline
   & \multicolumn {3} {c|}{$\pi\Sigma$} & \multicolumn {3} {c} {$\bar{K} N$} \\ 
  & \hfil $ a_{0} $ & \hfil  $ a_1 $  & \hfil   $ a_2 $  & \hfil   $a_0$   & \hfil   $a_1$  & \hfil  $a_2 $  \\
  \hline
  \centering A \ & \hfil $-2.27$ & \hfil $-2.29$ & \hfil $-1.50$ & \hfil $-1.83$ & \hfil $-1.95$ & \hfil $-1.60$ \\
  \centering B \ & \hfil $-2.27$ & \hfil $-2.29$ & \hfil $-2.05$ & \hfil $-1.83$ & \hfil $-1.95$ & \hfil $-2.31$ \\
  \centering C \ & \hfil $-$ & \hfil $-$ & \hfil $-1.96$ & \hfil $-$ & \hfil $-$ & \hfil $-1.96$ \\
\hline
\hline
\end{tabular}
\caption{\label{parameters} The subtraction constants, $a_0$, $a_1$ and $a_2$, in the loop functions, $G_0$, $G_1$ and $G_2$
without on-shell factorization, $A$ and $B$, and with on-shell factorization, $C$.}
\end{table}

We show the single-channel $\pi \Sigma$ and $\bar K N$ scattering amplitudes, respectively, in Figs.\ 2 and 3
and two diagonal scattering amplitudes of the $\pi \Sigma-\bar K N$ coupled channels, respectively, in Figs.\ 4 and 5,
where the scattering amplitudes are defined by
\begin{align}
  F_{\pi \Sigma} &= - \frac{M_\Sigma}{4\pi\sqrt{s}} T_{\pi \Sigma \, \pi \Sigma},  \\
  F_{\bar K N} &= - \frac{M_N}{4\pi\sqrt{s}} T_{\bar K N \, \bar K N}.
\end{align}
We also present the pole positions of the $T$-matrix for ${\bar K} N$ and $\pi \Sigma$ single-channel scatterings and
$\pi \Sigma - {\bar K} N$ coupled-channels in Table II and  Fig. 6.

Before explaining detailed results we would like to mention the following;
near the $\Lambda$(1405), a bound pole is found in the ${\bar K} N$ single channel, a resonance pole is found in the $\pi N$ single channel and two poles are found in the $\pi \Sigma-\bar K N$ coupled channels, in cases $A$ and $B$ as in case $C$.

\begin{figure}[htbp]
\begin{tabular}{cc}
      \begin{minipage}[t]{0.5\hsize}
        \centering
        \includegraphics[keepaspectratio,scale=0.6]{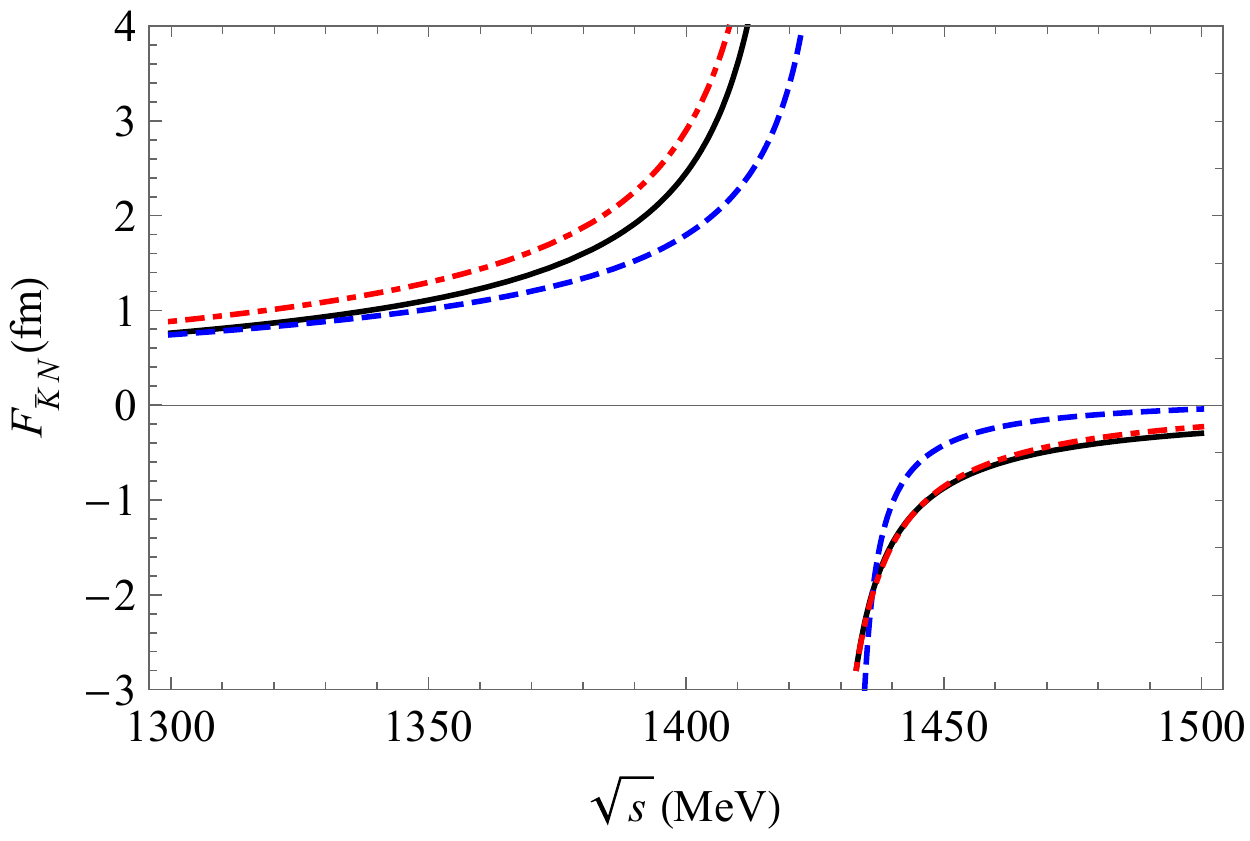}
      \end{minipage} &
      \begin{minipage}[t]{0.5\hsize}
        \centering
        \includegraphics[keepaspectratio,scale=0.6]{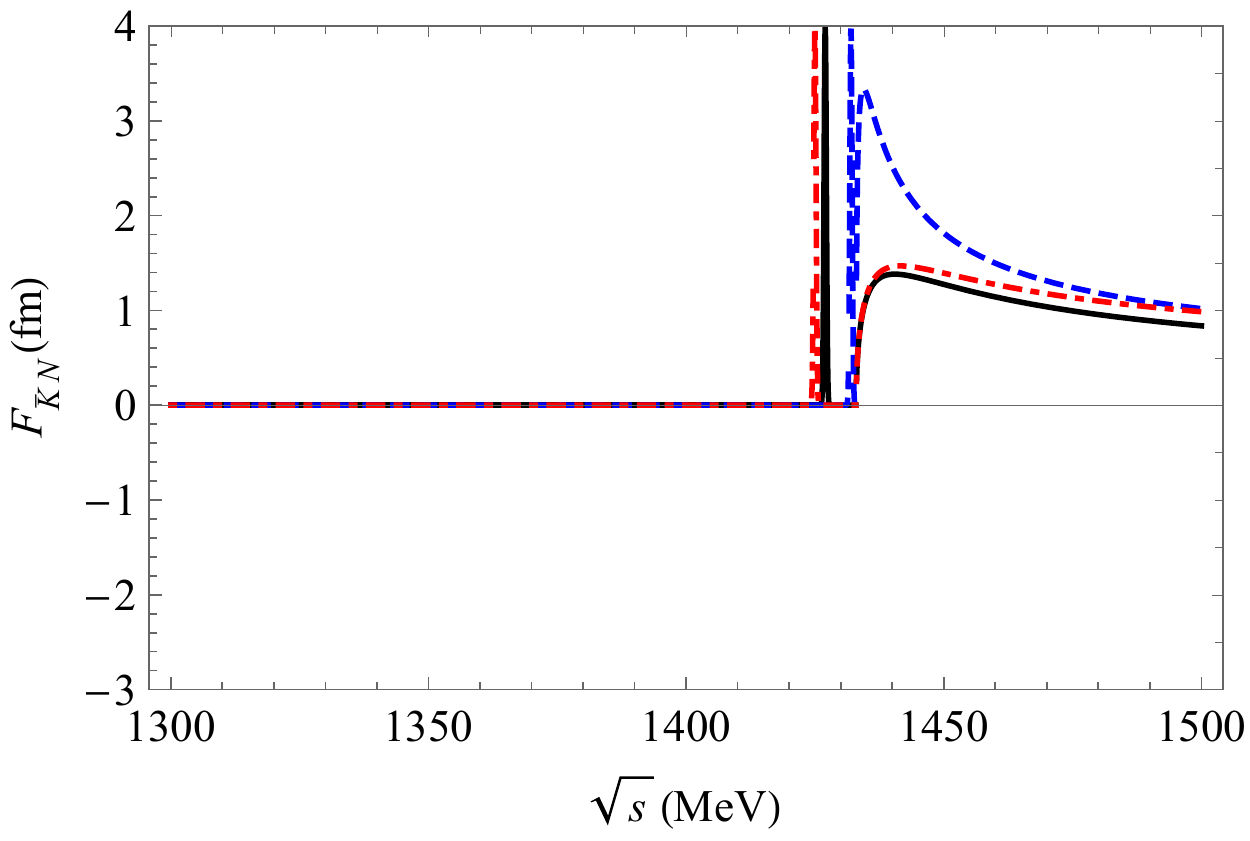}
      \end{minipage}
    \end{tabular}
\caption{(Color online) The real (left) and imaginary (right) parts of the scattering amplitude in the $\bar K N$ single channel, $F_{\bar K N}$.
The (blue) dashed lines, the (red) dot-dashed lines are the results without on-shell factorization, $A$ and $B$, respectively,
and the (black) solid lines are the results with on-shell factorization, $C$.}
\end{figure}

\begin{figure}[htbp]
\begin{tabular}{cc}
      \begin{minipage}[t]{0.5\hsize}
        \centering
        \includegraphics[keepaspectratio,scale=0.6]{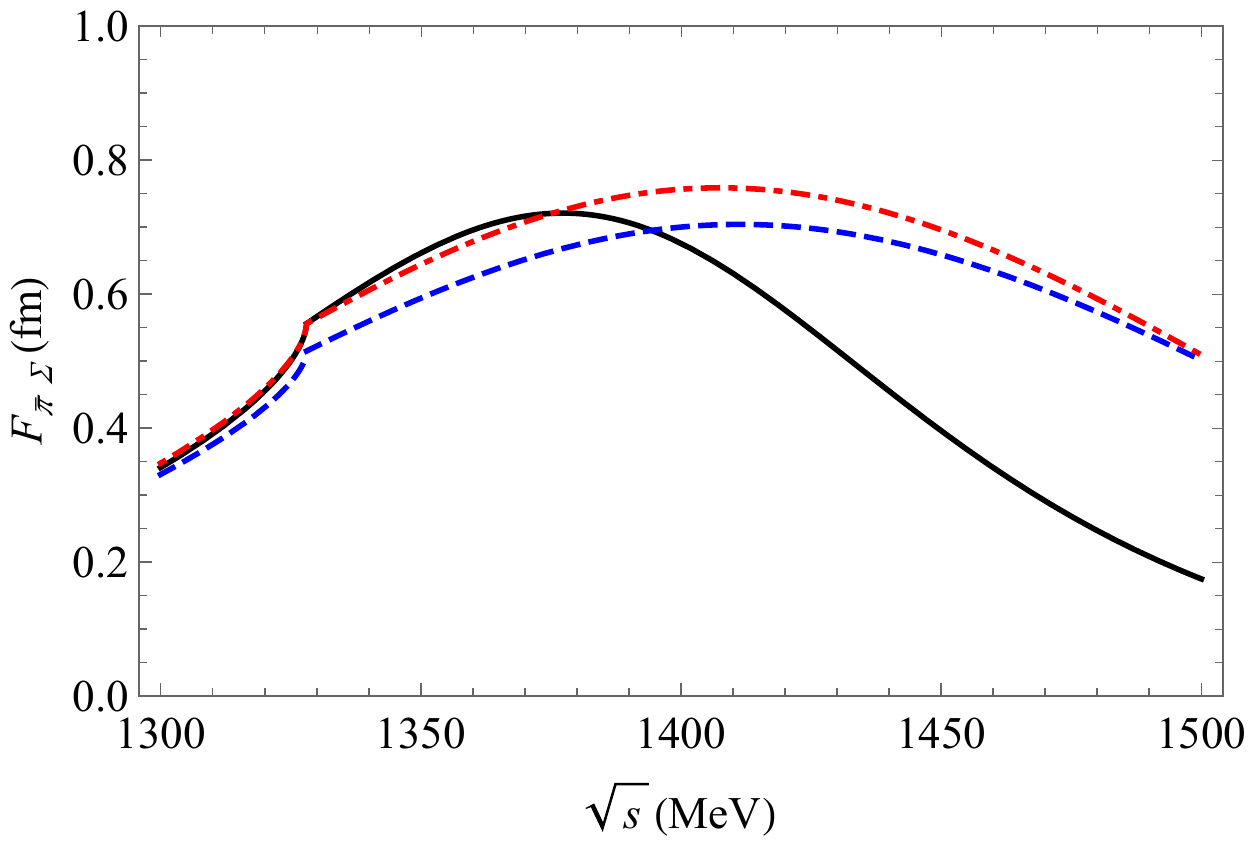}
      \end{minipage} &
      \begin{minipage}[t]{0.5\hsize}
        \centering
        \includegraphics[keepaspectratio,scale=0.6]{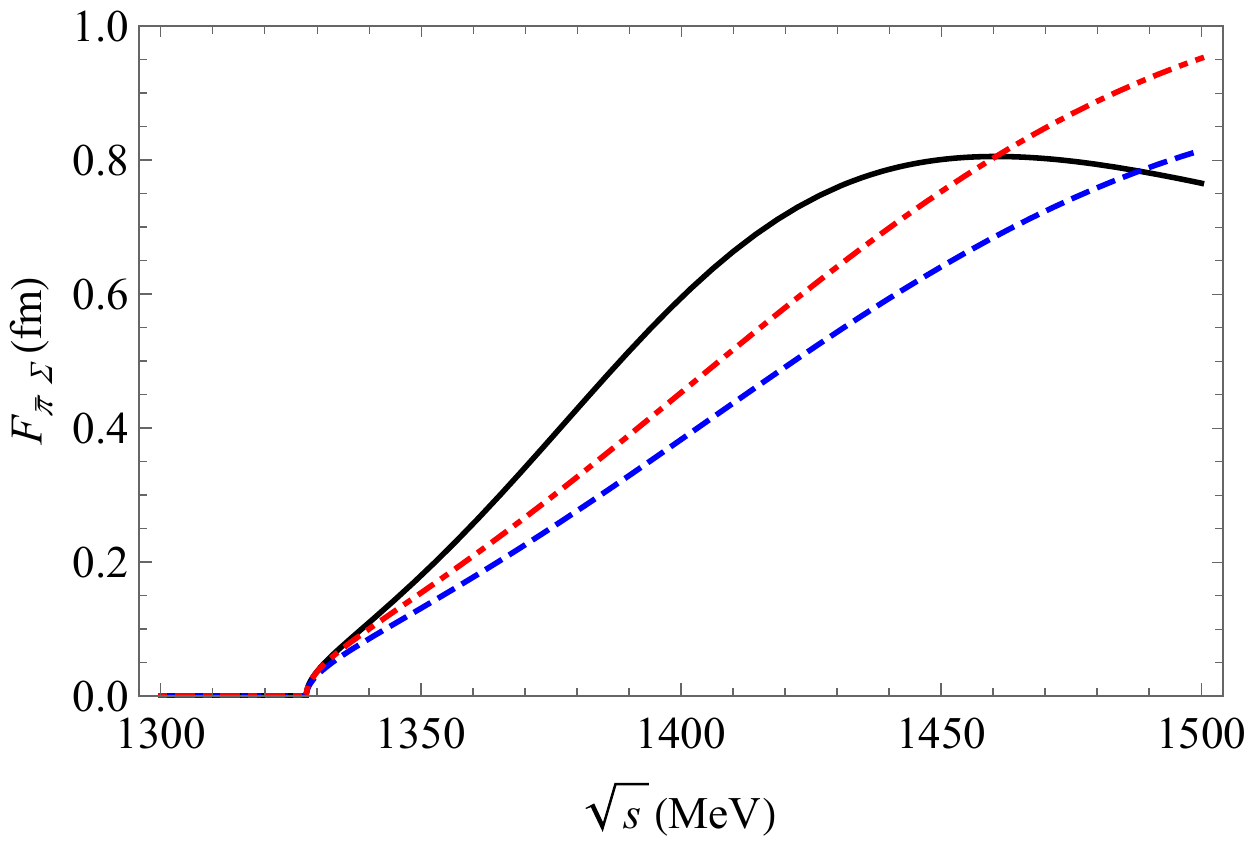}
      \end{minipage}
    \end{tabular}
\caption{(Color online) The real (left) and imaginary (right) parts of the scattering amplitude in the $\pi \Sigma$ single channel, $F_{\pi \Sigma}$.
The (blue) dashed lines, the (red) dot-dashed lines are the results without on-shell factorization, $A$ and $B$, respectively,
and the (black) solid lines are the results with on-shell factorization, $C$.}
\end{figure}

\begin{figure}[htbp]
\begin{tabular}{cc}
      \begin{minipage}[t]{0.5\hsize}
        \centering
        \includegraphics[keepaspectratio,scale=0.6]{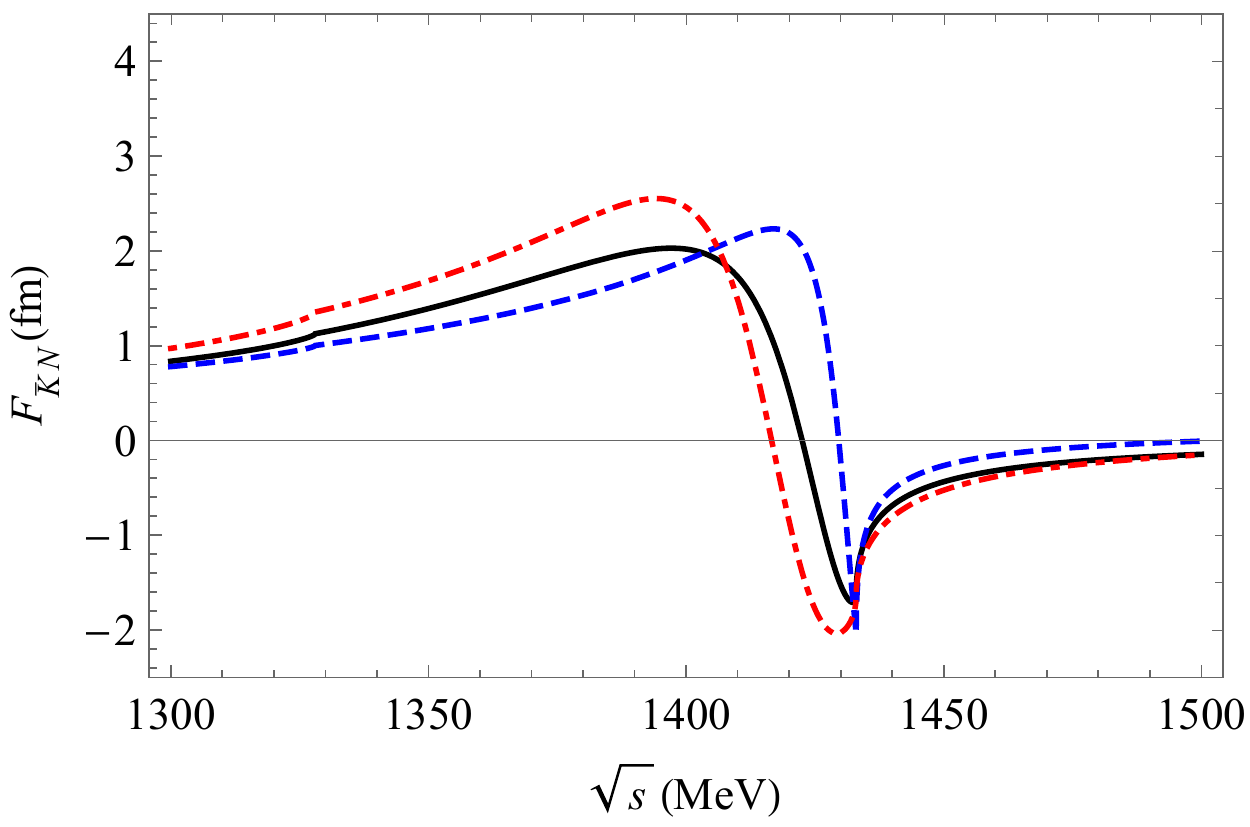}
      \end{minipage} &
      \begin{minipage}[t]{0.5\hsize}
        \centering
        \includegraphics[keepaspectratio,scale=0.6]{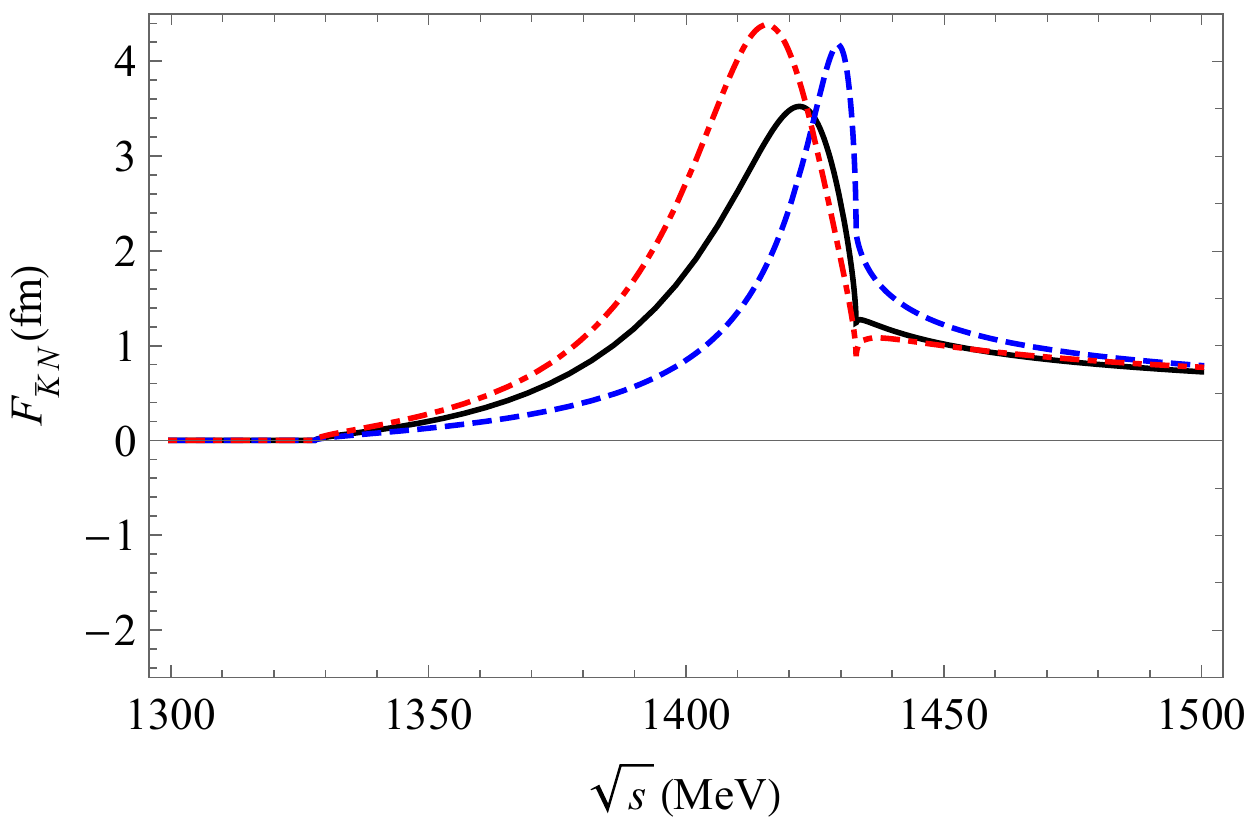}
      \end{minipage}
    \end{tabular}
\caption{(Color online) The real (left) and imaginary (right) parts of the scattering amplitude from $\bar K N$ to $\bar K N$ in the $\bar K N - \pi \Sigma$ coupled channels, $F_{\bar K N}$.
The (blue) dashed lines, the (red) dot-dashed lines are the results without on-shell factorization, $A$ and $B$, respectively,
and the (black) solid lines are the results with on-shell factorization, $C$.}
\end{figure}

\begin{figure}[htbp]
\begin{tabular}{cc}
      \begin{minipage}[t]{0.5\hsize}
        \centering
        \includegraphics[keepaspectratio,scale=0.6]{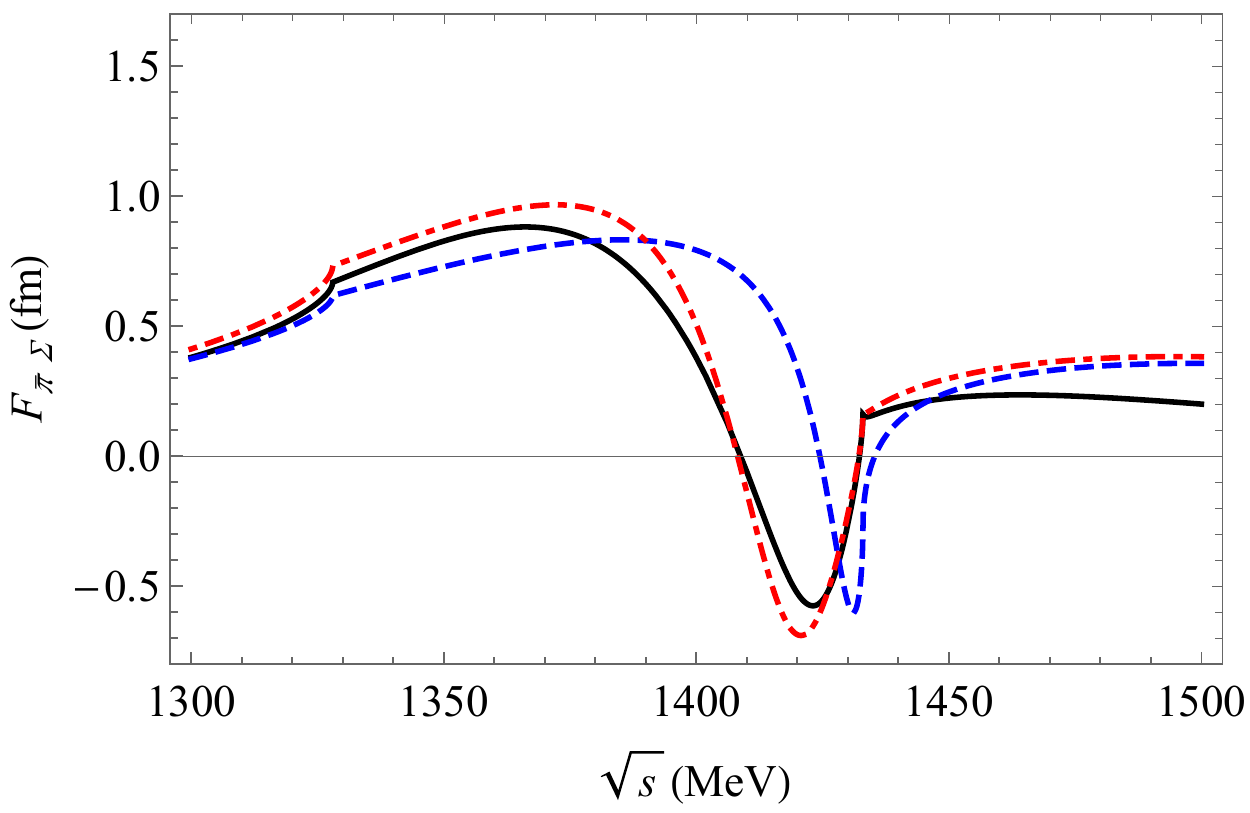}
      \end{minipage} &
      \begin{minipage}[t]{0.5\hsize}
        \centering
        \includegraphics[keepaspectratio,scale=0.6]{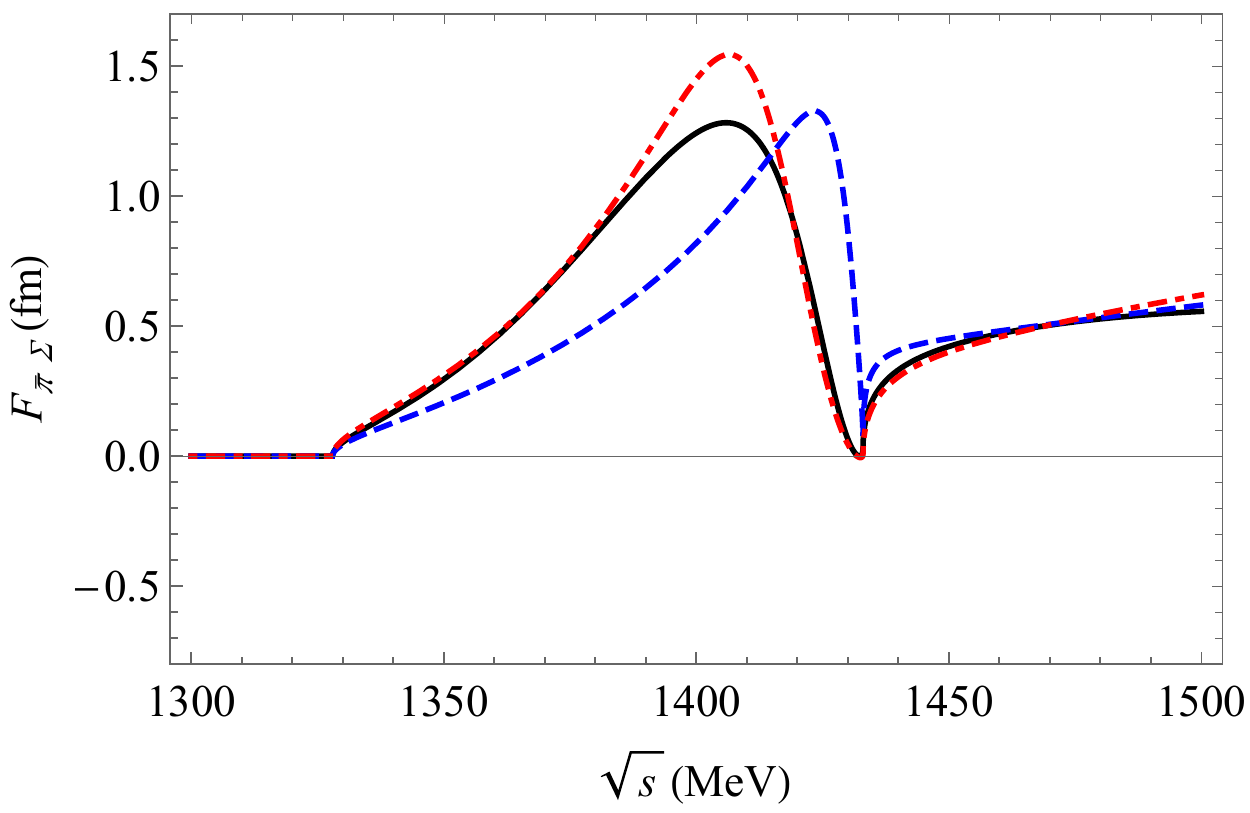}
      \end{minipage}
    \end{tabular}
\caption{(Color online) The real (left) and imaginary (right) parts of the scattering amplitude from $\pi \Sigma$ to $\pi \Sigma$ in the $\bar K N - \pi \Sigma$ coupled channels, $F_{\pi \Sigma}$.
The (blue) dashed lines, the (red) dot-dashed lines are the results without on-shell factorization, $A$ and $B$, respectively,
and the (black) solid lines are the results with on-shell factorization, $C$.}
\end{figure}

In the ${\bar K} N$ single channel, Fig.\ 3, the difference of the scattering amplitudes with and without on-shell factorization, $A$, $B$ and $C$, is small,
where the difference in the pole positions is also small.
Besides, the difference of $B$ and $C$ is smaller than that of $A$ and $C$.
These differences can be understood how far $\sqrt{s}$ is from the point where the renormalization condition is specified.
In the $\pi \Sigma - {\bar K} N$ coupled channels, Fig.\ 5, this tendency remains in the diagonal scattering amplitude, $F_{\bar K N}$, and the position of the pole close to that in the ${\bar K} N$ single channel.

In the $\pi \Sigma$ single channel, Fig.\ 4, the difference of the scattering amplitudes with and without on-shell factorization, $A$, $B$ and $C$, is small when $\sqrt{s} \lesssim 1400 \ {\rm MeV}$.
As $\sqrt{s}$ increases beyond $1400 \ {\rm MeV}$, the difference of the scattering amplitudes also increases.
The difference of the real part of the center-of-mass energy of the pole in the $\pi \Sigma$ single channel is also small but that of the imaginary part is considerably large:
the imaginary part in cases $A$ and $B$ is close to twice that in case $C$.
When the coupling between $\pi \Sigma$ and ${\bar K} N$ channels is turned on, Fig.\ 6, the behavior of the diagonal scattering amplitude, $F_{\bar K N}$, in the region $1400 \ {\rm MeV} \lesssim \sqrt{s} \lesssim 1500 \ {\rm MeV}$ is dominated by the pole close to that of the $\pi \Sigma$ single channel,
and therefore the difference of $A$, $B$ and $C$ becomes smaller.
The coupling, however, seems to enlarge the difference in the position of the second pole:
the real parts of the center-of-mass energies of the second pole in cases $A$ and $B$ differ from that in case $C$ about 20 ${\rm MeV}$ and the imaginary parts in cases $A$ and $B$ are twice as large as or even larger than twice that in $C$.
Again, the difference of $B$ and $C$ is smaller than that of $A$ and $C$, as expected.

\begin{table}[h]
\begin{tabular}{p{1.5cm}|p{3cm}p{3cm}|p{3cm}p{3cm}}
\hline
\hline
& \multicolumn{2} {c|} {single channel} &  \multicolumn{2} {c} {coupled channels} \\ 
& \hfil ${\bar K} N$ & \hfil $\pi \Sigma$ &  \multicolumn{2} {c} {${\bar K} N - \pi \Sigma$}  \\
\hline
\centering $A$ & \hfil $1432 \ \text{MeV}$ & \hfil $1388 - 179 i \ \text{MeV}$ & \hfil $1434 - 7 i \ \text{MeV}$ & \hfil $1418 - 160 i \ \text{MeV}$ \\
\centering $B$ & \hfil $1425 \ \text{MeV}$ & \hfil $1382 - 169 i \ \text{MeV}$ & \hfil $1419 - 19 i \text{MeV}$ & \hfil $1424 - 146 i \ \text{MeV}$ \\
\centering $C$ & \hfil $1427 \ \text{MeV}$ & \hfil $1388 - 96 i \ \text{MeV}$ & \hfil $1432 - 17 i  \text{MeV}$ & \hfil $1398 - 73 i \ \text{MeV}$ \\
\hline
\hline
\end{tabular}
    \caption{\label{fig:epsart} Pole positions of the $T$-matrix in the ${\bar K} N$ and $\pi \Sigma$ single-channel scatterings and the ${\bar K} N - \pi \Sigma$ coupled channels without on-shell factorization, $A$ and $B$, and with on-shell factorization, $C$.}
\end{table}

\begin{figure}[h]
  \begin{center}
        \includegraphics[keepaspectratio,scale=0.6]{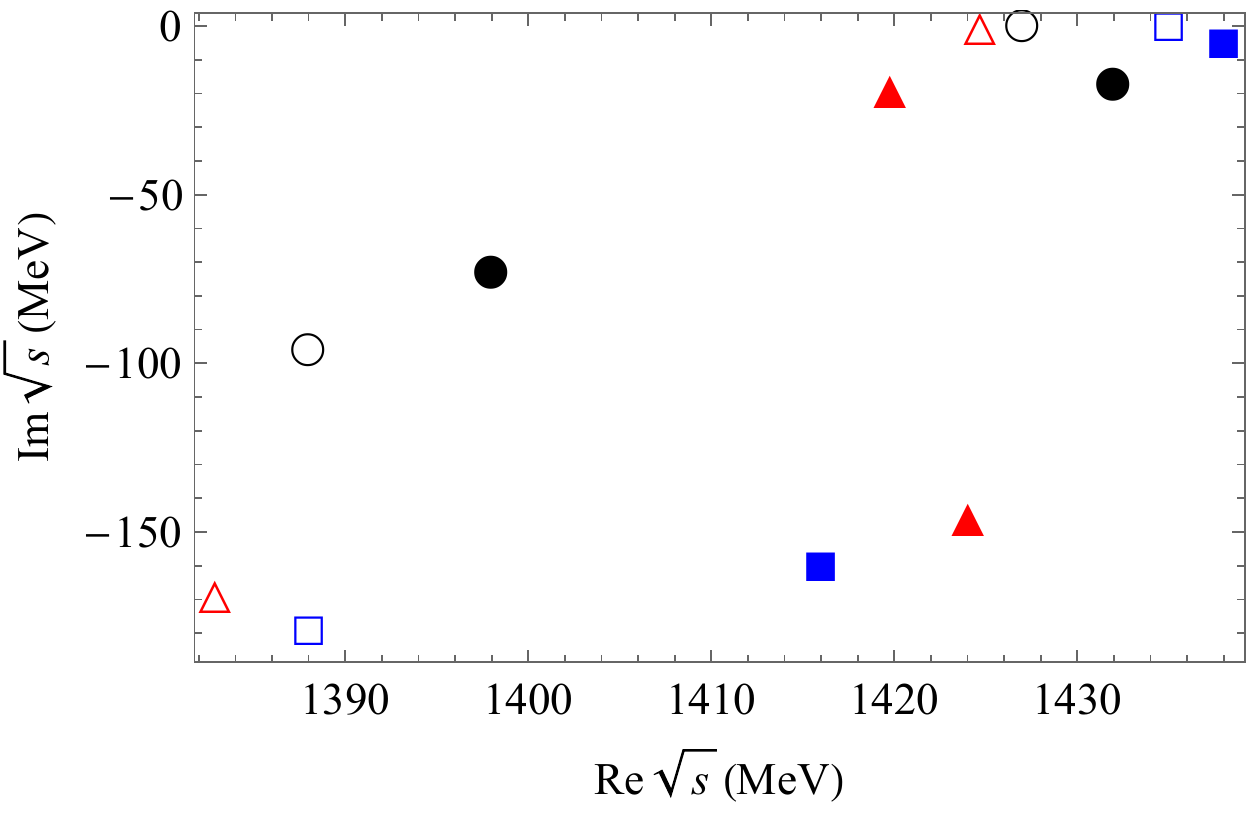}
    \caption{\label{fig:epsart} (Color online) Pole positions of the $T$-matrix in the ${\bar K} N$ and $\pi \Sigma$ single-channel scatterings (unfilled) and the ${\bar K} N - \pi \Sigma$ coupled channels (filled).
    The (blue) squares, (red) triangles are the results without on-shell factorization, $A$ and $B$, respectively,
    and the (black) circles are the results with on-shell factorization, $C$.}
  \end{center}
\end{figure}

Here, we discuss the origin of the contradiction about the second pole between the present work without on-shell factorization, the phenomenological approach without on-shell factorization and the chiral unitary approach with on-shell factorization.
In the present work, we regularize the divergent integrals by dimensional regularization and renormalize them by introducing counter terms,
where we impose renormalization conditions that the scattering $T$-matrix has the form of the Weinberg-Tomozawa interaction.
In the phenomenological approach without on-shell factorization, they regularize the divergent integrals by modifying the Weinberg-Tomozawa interaction to a separable potential with suitable cut-off functions.
Then, loop terms do not give infinite corrections to the tree term and are not renormalized.
However, loop terms do give finite corrections.
Thus, the physical scattering $T$-matrix is expected not to have the form of the Weinberg-Tomozawa interaction.
In the chiral unitary approach with on-shell factorization, the scattering $T$-matrix has the form of the Weinberg-Tomozawa interaction without renormalization,
as has already been mentioned.
This apparent unnecessity of renormalization in the chiral unitary approach with on-shell factorization seems to have caused confusions about the second pole between two approaches, the chiral unitary approach with on-shell factorization and the phenomenological approach without on-shell factorization.
In our opinion the origin of the contradiction about the second pole is whether the scattering $T$-matrix has the form of the Weinberg-Tomozawa interaction or not but not whether on-shell factorization is employed or not.
This does not necessarily mean, however, that the scattering $T$-matrix of the phenomenological approach is physically unreasonable.
This only means that the off-shell behavior of the $T$-matrix of the phenomenological approach is different from that of the Weinberg-Tomozawa interaction.
In fact, in Ref.\ {\cite{Revai:2017isg}} they have adjusted the potential so as to reproduce the available experimental data.

\section{Summary}
In this paper we studied unitarized chiral dynamics without on-shell factorization.
We showed that we can take a ladder sum of the Weinberg-Tomozawa interaction without on-shell factorization.
In the case of  coupled $n$-channels, the equation for the scattering $T$-matrix is a $2n$ by $2n$ matrix equation,
while it is an $n$ by $n$ matrix equation with on-shell factorization.
There appear three types of divergent loop functions while there is only one with on-shell factorization.
The divergent integrals are regularized by dimensional regularization and renormalized by counter terms.
Not only infinite but also finite renormalization is important in order for the renormalized physical scattering $T$-matrix to have the form of the Weinberg-Tomozawa interaction.
The scattering $T$-matrix without on-shell factorization has two poles in the complex center-of-mass energy plane as with on-shell factorization, the real part of which is close to the observed $\Lambda$(1405).
With and without on-shell factorization, the difference of the scattering $T$-matrix is small near the renormalization point, also close to the observed $\Lambda$(1405).
The difference, however, increases with the distance from the renormalization point.
In particular, the difference in the position of the second pole, close to the one in the $\pi \Sigma$ single channel, is considerably large, while that of the first pole, close to the one in the $\bar K N$ single channel, is small:
the imaginary part of the center-of-mass energy of the second pole without on-shell factorization is as large as or even larger than twice that with on-shell factorization.

Here, we summarize what should be done in near future.
\begin{itemize}
\item
The Klein-Gondon propagator should be replaced by the Dirac propagator for baryons.
\item
The $\pi \Sigma - {\bar K} N - \eta \Lambda - K \Xi$ coupled-channel calculation should be done.
\item
The results of the calculation should be compared with experiment.
\item
Application to other channels such as $S=-1$ and $I=1$ or $B(\text{baryon number}) = 2$ should be considered.
\end{itemize}

\section{Conclusion}
The conclusion of the present paper is as follows.

On the one hand, in the chiral unitary approach the calculation with on-shell factorization should be abandoned because it cannot be justified and the calculation without on-shell factorization is almost as easy as the calculation with on-shell factorization, i.e.\ to diagonalize matrices of $2n$ by $2n$ instead of $n$ by $n$.
On the other hand, in the phenomenological approach one should make sure that the meson-baryon $T$-matrix, not the meson-baryon potential, has the form of the Weinberg-Tomozawa interaction.

In the calculation without on-shell factorization the second pole does show up in the complex center-of-mass energy plane as with on-shell factorization.
Therefore, the appearance of the second pole is not the consequence of on-shell factorization.
However, the imaginary part of the second pole without on-shell factorization is as large as or even larger than twice that of the second pole with on-shell factorization,
which makes it hard to believe that the second pole plays an important role in the rather sharp resonant structure of $\Lambda(1405)$.
It is of doubt that the double-pole structure of $\Lambda(1405)$ is an inevitable conclusion of the Weinberg-Tomozawa interaction, i.e.\ chiral dynamics.
Clearly, thorough reanalysis of the problem without on-shell factorization would be necessary in order to draw the final conclusion.

After finishing this work we were informed the existence of Ref.\ \cite{Mai:2012dt}, in which the scattering $T$-matrix is calculated at next-to-leading order in the chiral expansion without on-shell factorization.
We could not have fully clarified the relation between the present work and Ref.\ \cite{Mai:2012dt}, because in which the explicit expressions for the scattering $T$-matrix are not given.
However, we have three types of meson-baryon loop functions while they give only one corresponding to $G_2$ in Eq.\ (5).
Therefore, there seem to be some differences in two works.

\begin{acknowledgments}
We would like to thank Yoshinori Akaishi, Akinobu Dote, Tetsuo Hyodo, Toru Sato, Koichi Yazaki and Toshimitsu Yamazaki for discussions.
\end{acknowledgments}


\begin{thebibliography}{99}
%\cite{Weinberg:1978kz}
\bibitem{Weinberg:1978kz} 
  S.~Weinberg,
  %``Phenomenological Lagrangians,''
  Physica A {\bf 96}, no. 1-2, 327 (1979).
%\cite{Gasser:1983yg}
\bibitem{Gasser:1983yg} 
  J.~Gasser and H.~Leutwyler,
  %``Chiral Perturbation Theory to One Loop,''
  Annals Phys.\  {\bf 158}, 142 (1984).
%\cite{Gasser:1984gg}
\bibitem{Gasser:1984gg} 
  J.~Gasser and H.~Leutwyler,
  %``Chiral Perturbation Theory: Expansions in the Mass of the Strange Quark,''
  Nucl.\ Phys.\ B {\bf 250}, 465 (1985).
%\cite{Scherer:2009bt}
\bibitem{Scherer:2009bt} 
  S.~Scherer,
  %``Chiral Perturbation Theory: Introduction and Recent Results in the One-Nucleon Sector,''
  Prog.\ Part.\ Nucl.\ Phys.\  {\bf 64}, 1 (2010)
%\cite{Kaiser:1995eg}
\bibitem{Kaiser:1995eg} 
  N.~Kaiser, P.~B.~Siegel and W.~Weise,
  %``Chiral dynamics and the low-energy kaon - nucleon interaction,''
  Nucl.\ Phys.\ A {\bf 594}, 325 (1995)
\bibitem{Oset:1997it} 
  E.~Oset and A.~Ramos,
  %``Nonperturbative chiral approach to s wave anti-K N interactions,''
  Nucl.\ Phys.\ A {\bf 635}, 99 (1998)
%\cite{Oller:2000fj}
\bibitem{Oller:2000fj} 
  J.~A.~Oller and U.~G.~Meissner,
  %``Chiral dynamics in the presence of bound states: Kaon nucleon interactions revisited,''
  Phys.\ Lett.\ B {\bf 500}, 263 (2001)
%\cite{Lutz:2001yb}
\bibitem{Lutz:2001yb} 
  M.~F.~M.~Lutz and E.~E.~Kolomeitsev,
  %``Relativistic chiral SU(3) symmetry, large N(c) sum rules and meson baryon scattering,''
  Nucl.\ Phys.\ A {\bf 700}, 193 (2002)
%\cite{Jido:2003cb} 
\bibitem{Jido:2003cb} 
  D.~Jido, J.~A.~Oller, E.~Oset, A.~Ramos and U.~G.~Meissner,
  %``Chiral dynamics of the two Lambda(1405) states,''
  Nucl.\ Phys.\ A {\bf 725}, 181 (2003)
%\cite{Magas:2005vu}
\bibitem{Magas:2005vu} 
  V.~K.~Magas, E.~Oset and A.~Ramos,
  %``Evidence for the two pole structure of the Lambda(1405) resonance,''
  Phys.\ Rev.\ Lett.\  {\bf 95}, 052301 (2005)
%\cite{Hyodo:2007jq}
\bibitem{Hyodo:2007jq} 
  T.~Hyodo and W.~Weise,
  %``Effective anti-K N interaction based on chiral SU(3) dynamics,''
  Phys.\ Rev.\ C {\bf 77}, 035204 (2008)
%\cite{Hyodo:2011ur}
\bibitem{Hyodo:2011ur} 
  T.~Hyodo and D.~Jido,
  %``The nature of the Lambda(1405) resonance in chiral dynamics,''
  Prog.\ Part.\ Nucl.\ Phys.\  {\bf 67}, 55 (2012)
%\cite{Ikeda:2011pi}
\bibitem{Ikeda:2011pi} 
  Y.~Ikeda, T.~Hyodo and W.~Weise,
  %``Improved constraints on chiral SU(3) dynamics from kaonic hydrogen,''
  Phys.\ Lett.\ B {\bf 706}, 63 (2011)
%\cite{Ikeda:2012au}
\bibitem{Ikeda:2012au} 
  Y.~Ikeda, T.~Hyodo and W.~Weise,
  %``Chiral SU(3) theory of antikaon-nucleon interactions with improved threshold constraints,''
  Nucl.\ Phys.\ A {\bf 881}, 98 (2012)
%\cite{Akaishi:2010wt}
\bibitem{Akaishi:2010wt} 
  Y.~Akaishi, T.~Yamazaki, M.~Obu and M.~Wada,
  %``Single-pole nature of Lambda (1405) and structure of K-pp,''
  Nucl.\ Phys.\ A {\bf 835}, 67 (2010)
%\cite{Revai:2017isg}
\bibitem{Revai:2017isg} 
  J.~R\'evai,
  %``Are the chiral based $\bar{K}N$ potentials really energy dependent?,''
  Few Body Syst.\  {\bf 59}, no. 4, 49 (2018)
%\cite{Myint:2018ypc}
\bibitem{Myint:2018ypc} 
  K.~S.~Myint, Y.~Akaishi, M.~Hassanvand and T.~Yamazaki,
  %``Single-pole Nature of the Detectable Lambda(1405),''
  PTEP {\bf 2018}, no. 7, 073D01 (2018)
%\cite{Mai:2012dt}
\bibitem{Mai:2012dt} 
  M.~Mai and U.~G.~Meissner,
  %``New insights into antikaon-nucleon scattering and the structure of the Lambda(1405),''
  Nucl.\ Phys.\ A {\bf 900}, 51  (2013)
\end{thebibliography}
\end{document}